\documentstyle[prl,twocolumn,aps,psfig]{revtex}

\def\epsfig#1#2#3#4
         {
         \epsfysize=#2 \vbox{ \hglue#3 \epsfbox[#4]{#1} }
         }
\def\epsfigrot#1#2#3#4
         {
         \epsfxsize=#2
         \setbox\rotbox=\hbox to #2{\epsfbox[#4]{#1}}
         \vbox{\hglue#3 \rotl\rotbox}
         }
\newbox\rotbox
\input rotate

\begin{document}
\title{Strong coupling resistivity in the Kondo model.}
\author{F. Lesage$^*$, H. Saleur$^{**}$.}
\address{$^*$ Centre de recherches math\'ematiques, Universit\'e
de Montr\'eal, C.P. 6128 Succ. Centre-ville, Montr\'eal, H3C-3J7.}
\address{$^{**}$Department of Physics, University of Southern California,
Los-Angeles, CA 90089-0484.}
\date{October 28, 1998.}
\maketitle

\begin{abstract}
By applying methods of integrable quantum field theory
to the Kondo problem, we develop a systematic perturbation
expansion near the IR (strong coupling) fixed point. This requires
the  knowledge  of an infinity of irrelevant operators and
their  couplings, which we all determine exactly. A 
low temperature expansion (ie all the corrections
to Fermi liquid theory) of the resistivity then follows, extending 
for instance the well known Nozi\`eres $T^2$ result in the 
exactly screened case to arbitrary order. The example of the ordinary Kondo model is worked out in details: we determine $\rho$ up to 
order $T^6$, and compare
the result with available numerical data. 
\end{abstract}

Theoretical progress in the understanding of the Kondo model
over the last twenty years has been remarkable (see \cite{Hewson} 
for a review). Most of the 
interest has focussed on the strong coupling fixed point, where
non perturbative techniques like the numerical renormalization group,
the Bethe ansatz \cite{oldba} and conformal field theory \cite{Ian}
have provided many powerful results. Yet, the
basic quantity that gave rise to the discovery of the Kondo physics
itself, the resistivity $\rho$, has  remained strangely resilient  to 
analytical understanding : although its behaviour is well understood, 
 based on various numerical renormalization
or approximation methods,  essentially no exact results
have been obtained for it - for instance, it cannot be computed
by the Bethe ansatz, in contrast with the specific heat, or the 
magnetic susceptibility. A notable exception to this state of affairs
is the result of Nozi\`eres \cite{Nozieres} for the ordinary Kondo problem, 
which gives the
 exact first order 
correction to the resistivity near the strong coupling (IR) fixed point (see below).
This result has since been generalized to the case of several channels and 
higher spins  \cite{Hewson}.

It is important here to emphasize that studying the vicinity of the strong 
coupling fixed point is a subtle challenge. In the case 
of the weak coupling (UV) fixed point, one at least knows the effective 
hamiltonian,
up to irrelevant terms, so the universal properties at high temperature
 can in principle be 
accessed perturbatively. As is well known, this  perturbation breaks down 
at some point.
A natural idea to study the low temperature properties
 is thus to perturb around the other, strong coupling fixed point. In that 
case however, all the operators are irrelevant. The leading irrelevant
operator is usually the only one kept in the computations: it is this
operator
that gives rise to the Fermi liquid or Nozi\`eres result in the exactly screened 
case,  or the results of Affleck and Ludwig in the overscreened case \cite{AL}.
To go beyond first order, one however needs to know the {\sl next} irrelevant 
operators,
together with their coupling constants.  This is a difficult problem, that cannot be answered  by qualitative arguments. 
As we will see below, even the form of the
operators is not obvious, and requires a complicated adjustment. 
The only way to determine precisely the hamiltonian near the IR fixed point
is to use, somehow, the fact that it is the ``large coupling limit'' of the 
known hamiltonian near the UV fixed point. How to do that in practice
however was a largely open problem up to now. A few years ago, Hewson \cite{HewsonI}
 introduced
a ``renormalized perturbation theory'' that should in principle answer the 
question : in practice however, the method seems very hard to
 implement,
and we are not aware of any quantitative 
correction to the Fermi liquid behaviour
obtained in that way. 

The main result of this paper is the exact determination of the hamiltonian
near the IR fixed point to {\sl all} orders. It is then a purely 
technical matter to determine the corrections to the resistivity. We 
carry them out to order $T^6$ in the ordinary Kondo problem, and compare 
with existing numerical data. 

Let us now define the problem in more details. We follow closely the 
notations of \cite{AL}. 
We take as the starting hamiltonian  
\begin{equation}
{\cal H}=\sum_{k,\mu} \psi^{\dagger\mu}_{k} 
\psi_{k\mu} \epsilon(k)+\lambda 
\vec S\cdot \sum_{k,k'}
\psi^{\dagger\mu}_{k} \frac{\vec\sigma_\mu^\nu}{2}
\psi_{k' \nu},
\end{equation}
with $\mu$ the spin label.  Assuming that the impurities are
dilute enough so  we can restrict to the case of only one impurity,
we  decompose as usual  the fermionic
field on spherical harmonics. Only the s-wave 
interacts with the impurity and, at low energies,
its field operator takes the form
\begin{equation}
\Psi(\vec r)=\frac{1}{2\sqrt{2}i\pi r}[e^{ip_F r}\psi_R(r)-
e^{-ip_F r}\psi_L(r)],
\end{equation}
with the proper linearisation of the dispersion relation
in that limit, ie $\epsilon_p\simeq v_F(p-p_F)\equiv 
v_F p'$.  
The resistivity can be obtained by using the Kubo formula
for the conductivity, following\cite{AL}
\begin{equation}
\frac{1}{\rho(T)}=\sigma(T)=2  \frac{e^2}{3m^2}\int 
\frac{d^3\vec p}{(2\pi)^3} \left[
-\frac{dn}{d\epsilon_k}\right] \vec p\cdot \vec p \ \tau(\epsilon_p)\label{kubo},
\end{equation}
with the single particle lifetime defined by
$1/\tau=-2 {\rm Im}\Sigma^R$.
$\Sigma^R$, the retarded self-energy, is obtained from the
three-dimensional Green's function which in turns follows from
the  one-dimensional s-wave contributions\cite{AL}
\begin{eqnarray}
G(\omega_M,\vec r_1,\vec r_2)-G^0(\omega_M,\vec r_1-
\vec r_2)= \nonumber\\
\frac{-1}{8\pi^2 r_1 r_2}  \left[
e^{-i k_F(r_1+r_2)} (G_{LR}^{(1D)}-G^{0(1D)}_{LR}) 
\right.
 \nonumber\\
\left. +e^{ik_F (r_1+r_2)} (G_{RL}^{(1D)}-G^{0(1D)}_{RL})
\right]. \label{eqi}
\end{eqnarray}
Here the superscript $0$ means that the Green's function is evaluated
with respect to the non-interacting
UV fixed point where $\psi_R(0)=\psi_L(0)$
as a consequence of the spherical harmonic decomposition.
Eq (\ref{eqi}) can be rewritten as
\begin{eqnarray}
G(\omega_M,\vec r_1,\vec r_2)-G^0(\omega_M,\vec r_1-
\vec r_2)= \\  \nonumber
G^0(\omega_M,\vec r_1)T(\omega_M)G^0(\omega_M,-\vec r_2).
\end{eqnarray}
One can then sum over the dilute array of random 
impurities, restoring  translation invariance.  The summation
over multi-impurity terms leads to the  self energy
$\Sigma(\omega_M)=n_i T(\omega_M)$, to
first order in $n_i$, the impurity
density. As usual, impurity-impurity interactions are neglected, and 
the following results 
should be reliable for dilute impurities (see \cite{AL}, \cite{IanI} for
more discussion on that point).
The retarded self-energy is found by analytical continuation,
$i\omega_M\rightarrow \omega+i0$. It leads to the single particle
lifetime, and thus to the resistivity using (\ref{kubo}). 

It is thus clear, as stressed in \cite{AL},
 that the $RL$ and $LR$ components of the 
$1D$ Green's functions are the only necessary ingredients to compute
the resistivity, and we will concentrate on these in the following.
At the IR fixed point, the spin is completely screened and the 
interaction with the impurity amounts to a phase
shift, which in the one-channel case is simply encoded in the
boundary conditions $\psi_R(0)=-\psi_L(0)$: this leads to the result
$\Sigma^R(\omega)=-\frac{in_i}{\pi \nu}$, where $\nu$ is the density of states 
per spin. 

To proceed, we bosonize the one dimensional theory, setting $\psi_{R/L,\mu}=
\exp\left(\pm i\sqrt{4\pi} \phi_\mu\right)$ (our normalization for the fermions
follows \cite{AL} anddoes not contain the usual factors of $2\pi$). The spin and
charge degrees of freedom then completely decouple from each other
and, using spin and charge fields
\begin{equation}
\phi_s=\frac{1}{\sqrt{2}}(\phi_\uparrow-\phi_\downarrow), \
\phi_c=\frac{1}{\sqrt{2}}(\phi_\uparrow+\phi_\downarrow),
\end{equation}
the interaction only involves the spin field, with hamiltonian
\begin{equation}
{\cal H}_{UV}={\cal H}^*+
\lambda[S_- e^{i\sqrt{2\pi}\phi_s(0)}+c.c.]\label{boshamil}
\end{equation}
where right and left movers have been combined in one
field, and ${\cal H}^*=\frac{1}{2}\int_{-\infty}^0
dx [(\partial_x\phi_s)^2+\Pi_s^2]$ is a free boson hamiltonian.
To write (\ref{boshamil}) we have performed an additional 
canonical transformation to get rid of the $S_z$ terms.
In the usual approximation where  the band structure
does not  mix up spin and charge, the RG flow 
is completely controlled by the spin sector.

The hamiltonian (\ref{boshamil}) defines an integrable boundary quantum 
field theory \cite{GZ}. Although much progress has been made in 
computing physical properties by using integrability \cite{oldba},
including some correlators \cite{SLS},
the resistivity so far has not been accessible. The two reasons for that
are: the divergence of the form factors expansion when the perturbing operator
has dimension 1 as in (\ref{boshamil}), and the fact that the resistivity is really
a three dimensional quantity, following in a very non trivial way only from
the one dimensional computations. 

Our strategy here will be to use
integrability to determine the exact form of the IR hamiltonian. Once 
this is obtained, we will be able to
do perturbation theory near the IR fixed point: short of a Bethe ansatz 
determination of the resistivity,  this seems the 
most powerful approach at the present time.

We write 
\begin{equation}
{\cal H}_{IR}\equiv {\cal H}^*+\delta{\cal H}={\cal H}^*+
\sum_{k=0}^\infty u_{2k+1} {\cal O}_{2k+2}(0)
\label{irhamil},
\end{equation}
where the coupling constants $u$ and operators ${\cal O}$ have 
 to be determined. In (\ref{irhamil}), ${\cal H}^*$ is still 
a fre boson hamiltonian. However, due to the modified boundary condition
$\psi_L=-\psi_R$ at the IR fixed point, we have $\phi_L(0)=\phi_R(0)+i
\sqrt{\pi\over 4}$. 
 
It is fair to say at this stage that the determination of (\ref{irhamil}) 
 follows implicitely from the works of Bazhanov
 et al. \cite{BLZ}. We present here a much more direct, 
and generalizable, argument.
The main idea is the following: we first restrict to $T=0$,
and we switch the description from ``open string channel'' as in (\ref{boshamil})
(where imaginary time  is in the $y$ direction), to the ``closed string channel'',
where imaginary time is  now along $x$. In that case, the theory is
 described by a free 
boson hamiltonian, and all the interactions are encoded in a 
boundary state $|B\rangle$. The latter takes a simple form \cite{GZ}
because of integrability:
\begin{equation}
|B\rangle\propto\exp \left[ \int \frac{d\beta}{2\pi}
K(\beta-\log T_B)
\sum_{\epsilon=\pm} Z^*_{L\epsilon}(\beta)
Z^*_{R\epsilon}(\beta)
\right] |0\rangle.
\end{equation}
Here, $|0\rangle$ is the ground state of the theory in the open
string channel, $T_B$ is proportional to the Kondo temperature (we discuss 
their exact relation below),
 the $Z^*_\epsilon$ are creation operators for
the kinks and antikinks of the massless sine-Gordon theory \cite{FSW} - that is,
 the basis
diagonalising the interaction. $\beta$ is the rapidity of the massless kinks,
satisfying $e=\pm p=e^\beta$ for right (resp. left) movers.
We do not need the
precise way in which the kinks and antikinks behave to explain 
the argument: it suffices to say that their scattering in 
the bulk is factorized,
and that, in the open channel picture, they bounce on the boundary in a factorized way too,
with a reflection matrix that is simply related with $K$.  
Here, $K(\beta)=-i\tanh\frac{\beta}{2}$. To proceed, 
let us now expand $|B\rangle$ around the IR fixed point,
ie for $T_B$ large. We get
\begin{equation}
|B\rangle=\exp\left[-\sum_{k=0}^\infty \frac{1}{(2k+1)T_B^{2k+1}}
\widehat{{\cal I}}_{2k+1}\right]|B\rangle^*
\end{equation}
where $|B\rangle^*$ is the IR boundary state and
the $\widehat{{\cal I}}_{2k+1}$ are the standard  conserved 
quantities of the theory:
they act diagonally on the multiparticles states 
with eigenvalues
\begin{equation}
\widehat{{\cal I}}_{2k+1}|\beta_1...\beta_n\rangle_{C_1,\epsilon_1\cdots}=
 \left(\sum_i e^{(2k+1)\beta_i}
\right)
|\beta_1...\beta_n\rangle_{C_1,\epsilon_1,\cdots},
\end{equation}
with $C_i$ the chirality. By comparing with 
the general formula  
$|B\rangle={\cal P}\exp[-\int dy \delta {\cal H}]|B\rangle^*$, where ${\cal P}
\exp$ is the y-path ordered exponential, 
we can identify the hamiltonian. The last step to do so
is to reexpress the $\widehat{{\cal I}}_{2k+1}$ in terms of (integrals of) 
local operators
of the theory. This is easy to do, using the condition that these operators
must be mutually commuting  (how to find  the proper 
normalization will be discussed in a subsequent publication \cite{LS}).
After a Wick rotation to go back to the open string channel, 
the final result is 
\begin{equation}
u_{2k+1}=\frac{1}{\pi^k (k+\frac{1}{2}) (k+1)!}
T_B^{-(2k+1)}
\end{equation}
together with the form of the operators \cite{CQ}
\begin{eqnarray}
{\cal O}_2&=&-{1\over 4\pi}\left(T_{ww}+T_{\bar{w}\bar{w}}\right)\nonumber\\
{\cal O}_4&=&{1\over 4\pi}\left(:T_{ww}^2:+w\to\bar{w}\right)\nonumber \\
{\cal O}_6&=& -{1\over 4\pi}\left(:T_{ww}^3:+{1\over 4}
 :T_{ww}\partial_w^2T_{ww}:+w\to\bar{w}\right)\\
&\ldots\nonumber
\end{eqnarray}
Here, the basic object is the stress 
energy tensor $T_{ww}=-2\pi :\left(\partial_w\phi_s\right)^2:$ ($w=-y+ix$); it is 
the leading irrelevant operator near the IR fixed point, a result  well
known from previous work \cite{AL}. Up to an overall normalization, there is 
no ambiguity it: it is the only operator of dimension 2, up to
total derivatives. This is hardly true however for the next to leading 
irrelevant operators: they involve very particular combinations of 
powers of derivatives of $\phi_s$, 
and could not have been guessed on general grounds, without using the
constraint of integrability. The case of ${\cal O}_2$ is an exception,
where, after refermionization, one finds
\begin{equation}
{\cal O}_2\propto  2 :\psi^\dagger_\uparrow \psi_\uparrow
\psi^\dagger_\downarrow \psi_\downarrow: -i
:\psi^\dagger_\uparrow \partial_y \psi_\uparrow :
-i:\psi^\dagger_\downarrow \partial_y \psi_\downarrow : 
\end{equation}
It is  proportionnal to
$\vec J(0)\cdot \vec J(0)$ (up to total derivatives), 
with $\vec J$ the spin current \cite{Iannpb}.

Notice  that the more irrelevant operators in (\ref{irhamil})
 come with prefactors that
are inverse powers of the Kondo temperature: due to this rescaling,
they {\sl cannot} be neglected, and a perturbation expansion to $n^{th}$ non trivial
order near the IR fixed point requires the knowledge of 
the terms in (\ref{irhamil}) up to $k=n$. This is in contrast with the situation near
the UV fixed point, where the coefficients of the irrelevant operators
are dictated by the microscopic theory, and usually all of the same order, so only
the leading relevant one remains in the universal scaling regime. Here,
the vicinity of the IR fixed point is fully determined by the constraint
that it is the large coupling limit of the UV one, and this 
is what produces the fine tuning.

It is important to emphasize that our IR hamiltonian makes sense only 
within a well defined regularization scheme. Since all this is based
on the formalism of integrable quantum field theories
where no cut-off ever appears,  we have to 
 regularize our integrals by contour splitting, as  explained below.

Let us now discuss how to compute the resistivity. 
The first operator perturbing the 
fixed point is $:(\partial_y \phi_s)^2:$, the energy 
momentum tensor.  To order ${1\over T_B}$ for instance,
the correction to the $RL$ 1D Green's function for
the up-spin electrons reads
\begin{eqnarray}
\delta G_{RL}^{(1)}=\frac{2}{T_B}\int_{-\beta/2}^{\beta/2} dy
 e^{i\omega_M y}
\int_{-\beta/2}^{\beta/2} dy' 
\label{firstorder}\\ \nonumber
\langle e^{i\sqrt{2\pi}(\phi_c+\phi_s)} :(\partial_{y'}\phi_s)^2:
e^{-i\sqrt{2\pi}(\phi_c+\phi_c)}\rangle^*,
\end{eqnarray}
where the propagators have to be evaluated with the IR fixed point 
hamiltonian. The integrals in (\ref{firstorder}) can be quickly evaluated
by closing the contour in the complex plane using the periodicity of the integrand. The result is simply
\begin{equation}
-\frac{2i\pi}{T_B}
 e^{-\omega_M (r_1+r_2)} \omega_M \epsilon(\omega_M).
\end{equation}
Since this term is purely imaginary, it contributes a real term 
to the self energy, and  does not affect the resistivity. 
At order ${1\over T_B^2}$, ${\cal O}_2$ is still
the only operator to contribute. The integrals now present divergences where 
two of the ${\cal O}_2$ operators come close. These divergences are simply
regulated by shifting one of the contours slighlty off in the imaginary 
direction. One finds finally
\begin{equation}
\Sigma^R(\omega)=-\frac{in_i}{2\pi \nu}
\left[ 2+i\frac{\omega}{T_B}-\frac{1}{4T_B^2}(
3\omega^2+(\pi T)^2)\right],
\end{equation}
in agreement with \cite{AL}. 

We now extend the computation beyond this order. 
At  order $1/T_B^3$, it becomes necessary to include,
in addition to  the energy momentum tensor,  the
higher conserved operator ${\cal O}_4$. In manipulating ${\cal O}_4$, one has
 to be very careful that it is not a primary field: its correlation 
functions on the cylinder (ie at finite temperature) contain anomalous terms,
that can be computed from general conformal field theory arguments (for instance,
one finds $:T_{ww}^2:=4\pi^2 :(\partial_w\phi_s)^4:-2\pi :\partial_w\phi_s
\partial_w^3\phi_s:-2\pi^3T^2 :(\partial\phi_s)^2:+{3\pi^4T^4\over 20}$). 
As before, this order does
not contribute to the resistivity.  At  
order ${1\over T_B^4}$,  ${\cal O}_2$ and ${\cal O}_4$ are still the only operators 
that contribute, and we find
\begin{eqnarray}
&\Sigma^R(\omega)=-\frac{in_i}{2\pi \nu}
\left[ 2+i\frac{\omega}{T_B}-\frac{1}{4T_B^2}\left(
3\omega^2+(\pi T)^2\right)\right. \nonumber \\ \nonumber
 &-i\left(\frac{5}{12}+\frac{3}{24 \pi}\right)\left(\frac{\omega}{T_B}\right)^3
-i\left(\frac{1}{4}+\frac{1}{8\pi}\right) \frac{\omega}{T_B}
\left(\frac{\pi T}{T_B}\right)^2 \\
\nonumber
&+\left(\frac{35}{192}+\frac{7}{32\pi}\right)
\left(\frac{\omega}{T_B}\right)^4+
\left(\frac{19}{96}+\frac{5}{16\pi}\right)
\left(\frac{\pi T}{T_B}\right)^2 
\left(\frac{\omega}{T_B}\right)^2
\\ & \left.+
\left(\frac{11}{192}+\frac{3}{32 \pi}\right)
\left(\frac{\pi T}{T_B}\right)^4 \right]
\end{eqnarray}
In these computations, very strong 
short distance divergences are of course generally 
encountered (there are no large distance 
divergences due to the   temperature $T$
that acts as a cut-off). 
To regularize them, as for the second order case, 
we slightly move the contours off one another, and use the residue theorem
to evaluate the resulting integrals. The commutativity of all the 
$\widehat{{\cal I}}_{2k+1}$ means that the  short distance expansions
of the operators  ${\cal O}_{2k+2}$
have simple poles   whose residues are total derivatives: this 
ensures that the final result does not depend on the order in which this 
regularization is implemented. 

We have carried out the computation further
to fifth and sixth order. The expression for the self energy is too
bulky to be given here, but, by using the   Kubo formula as in \cite{AL},
it leads to our main result for the resistivity
\begin{eqnarray}
\rho(T)&=\frac{3n_i}{\pi (ev_F \nu)^2} \left[
1-\frac{1}{4}(\frac{\pi T}{T_B})^2 
+\left(\frac{13}{240}+\frac{3}{20\pi}\right) 
(\frac{\pi T}{T_B})^4+\right. \nonumber \\ 
&\left .\left(
\frac{47}{10080}-\frac{1}{8\pi}-\frac{53}{336\pi^2}
\right)(\frac{\pi T}{T_B})^6 +\cdots\right]\label{main}
\end{eqnarray}
where $\nu$ is the density of states per spin. The order $T^2$ allows
us to quickly match normalizations with other 
works - see for instance appendix K in the book by Hewson \cite{Hewson}
and we find $T_B={2T_K\over \pi w}$, $T_K$ the usual Kondo
temperature, $w=0.41071$ the Wilson number (compared to the review by Andrei et al. \cite{oldba}, $T_B=2T_0$).
The orders $T^4$ and $T^6$ in (\ref{main}) are new.

In general theories, one  
expects  IR expansions to be only asymptotic. However, for integrable quantum impurity 
problems, these expansions often turn out to be  convergent.  Assessing the 
status of the  $T^2$ expansion of the resistivity is a bit difficult: 
based on our numerical values, it seems to converge, but  with a 
disappointingly   small radius of convergence. That the radius is so small
presumably explains why estimates of $T_K$ from numerical or experimental
data on the resistivity often differ significantly from estimates based eg 
on the susceptibility (where convergence is good, as is easily 
checked on the  Bethe ansatz results). The comparison of 
our formula (\ref{main}) with the result
of  numerical renormalisation group methods
carried out by Costi et al. \cite{costi} on the Anderson model is shown on the 
following figure.  Clearly  more orders would be needed to 
go beyond $T/T_B\approx .2$. Pad\'e approximants however turn out to be quite 
stable, and agree well with the data up to $T/T_B\approx 1$, which is well in the crossover
regime. We thus believe that our results present a rather complete analytical understanding of the strong coupling resistivity. 
\begin{figure}[tbh]
\centerline{\psfig{figure=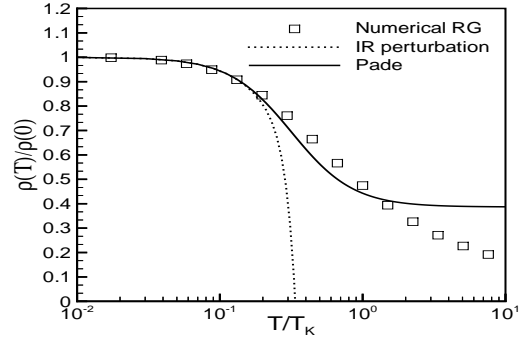,height=2.0in,width=3.0in}}
\caption{Comparison of IR perturbation theory with numerical RG results.}
\end{figure}

\smallskip
\noindent{\sl Acknowledgments}: this work was supported by the DOE and the NSF (through
the NYI program). We thank I. Affleck, T. Costi and A. Hewson
for discussions. We also thank 
 T. Costi for kindly providing his RG results\cite{costi}.

\end{document}